\documentclass{PoS}

\usepackage{bbold}
\usepackage{graphicx}

\newcommand{\be}{\begin{equation}}
\newcommand{\ee}{\end{equation}}
\newcommand{\beq}{\begin{eqnarray}}
\newcommand{\eeq}{\end{eqnarray}}
\newcommand{\usual}{1-(n-m+1) \frac{q}{3D} \frac{\partial D}{\partial q}}

\title{The probability of inflation in Loop Quantum Cosmology}

\ShortTitle{Probability of inflation in LQC}

\author{\speaker{William Nelson}%
        \thanks{This work was done in collaboration with Cristiano Germani 
                and Mairi Sakellariadou~\cite{us}}
        King's College London\\
        E-mail: \email{william.nelson@kcl.ac.uk}}

\abstract{The probability of there being sufficient inflation to solve the
fine-tuning associated with the horizon and flatness problems has recently been
shown to be exponentially small, within the context of classical general relativity.
Here this result is extended by considering loop quantum gravity effects, that 
are significant at small scales. In addition to accounting for high-energy 
departures from classicality, it is shown that, in contrast to the classical case, the
loop quantum cosmological probability measure is naturally finite, at least in some 
well defined region. It is also shown that these loop quantum gravity corrections can 
overcome the classical suppression of the probability only for extremely unnatural
choices of ambiguity parameters, implying that single field, slow-roll inflation
is exponentially unlikely.}

\FullConference{From Quantum to Emergent Gravity: Theory and Phenomenology\\
		 June 11-15 2007\\
		 Trieste, Italy}

\begin{document}

\section{Introduction}
Cosmological inflation is the most popular method of explaining the fine-tuning
difficulties associated with the standard hot big bang model~\cite{infl1} 
(although many other models exist~\cite{stringgas,cyclic,sling}). Inflation
is essentially an accelerated era of expansion that took place in the early 
universe, and it is easily motivated by the basic principles of general relativity
and quantum field theory. Despite its many successes, inflation can only be said to
solve the fine tuning problem if its onset is independent of the initial conditions.
However the fact that a quantum theory of gravity has yet to be fully developed means
that even when this issue is addressed, robust conclusions cannot be drawn~\cite{piran,
calzettamairi}. This issue of initial conditions can be overcome by using a measure
based on the dynamical structure of the theory and assuming some late time equivalence
between distinct universes. This was the method used in Ref.~\cite{GT}, where it
was shown that the probability of a universe at the end of an inflationary phase having
had $N$ or more efoldings of single field, slow-roll inflation is proportional to
$\exp \left( -3N\right)$.  Since $N$ is typically required to be approximately $60$, this would imply
that the onset of inflation needs to be significantly fine-tuned, however the calculation
is dependent on the equations of classical general relativity being valid at all scales.
In addition there is a restriction on types of inflationary potential that are amenable to
this approach, that was not discussed by the authors of Ref.~\cite{GT}.

Here the effect of the quantum gravity in the early universe will be accounted for
by considering corrections arising from loop quantum cosmology. The probability
of having $N$ or more efoldings of inflation within this setting will be investigated
and compared to the classical result and precise constraints on the validity
of this method will be discussed. A more detailed derivation and discussion can be found
in Ref.~\cite{us}.

\section{Loop Quantum Cosmology}
Loop quantum gravity~\cite{rovelli,tie} is a background independent, non-perturbative
quantisation of general relativity, which has been shown to be well behaved at
classical singularities such as the big bang. Despite its successes the full theory,
in an inhomogeneous setting is not entirely understood and in many cases it is not even 
possible to define a continuum limit. By restricting the symmetries of the theory
however, it is possible to produce a well defined classical limit. In particular
we are interested in homogeneous, isotropic cosmologies, which make the theory
tractable. To help ensure that the symmetry reduction doesn't induce additional effects,
wherever possible, the derivation follows that of the full theory~\cite{martin} (see also
Martin Bojowald's contribution to these proceedings). A brief introduction to LQC is given
below and follows the notation given in~\cite{us2,us3}. For a more complete description 
of the formalism see~\cite{ABL}.

Loop quantum cosmology (LQC) formulates general relativity in terms of Ashtekar variables:
$SU(2)$ valued triads $E_{\rm i}^a$ and a $SU(2)$ connection $A_{\rm i}^a$ (where 
${\rm ijk}$ etc. are $SU(2)$ indices, whilst $abc$ etc. are coordinate indices). The 
quantisation procedure uses holonomies of $A_{\rm i}^a$ along a specified edge $e$,
\be
 h_e(A) = {\cal P} \exp \int_e \dot{\gamma}^a(s) A^{\rm i}_a
\left(\gamma(s) \right)\tau_{\rm i} ds~,
\ee
where ${\cal P}$ infers path ordering on the exponential, $\dot{\gamma}^a$
is the tangent vector along the edge $e$, and $\tau_i$ are the basis of the $SU(2)$ Lie
algebra. Restricting to isotropic, homogeneous systems means we need only consider straight
edges along the integral curves of the basis vectors $X^a_{\rm i}$.
In this case the connection is given by a (dynamic)
multiple of the basis one forms $\omega_a^{\rm i}$, $A^{\rm i}_a = \tilde{c}(t) \omega^{
\rm i}_a$, whilst the triad is $E^a_{\rm i} = \sqrt{^0g} \tilde{p}(t) X^a_{\rm i}$, with
$^0g$ is the determinant of the fiducial metric\footnote{ This fiducial
metric is a complication that arises only for open universe and is used to define
the volume to which integral are restricted to ensure they remain finite. Clearly 
physical results must not depend on this volume\cite{Vandersloot_PhD}.}. With this
the holonomies becomes simply,
\beq
 h_{\rm i}(A)  &=& \exp \left[ \frac{-i\mu_0\sigma_{\rm i}}{2}  \tilde{c} 
\right]~,  \nonumber \\
&=& \cos \left( \frac{\mu_0 \tilde{c}}{2} \right) +2\tau_{\rm i} \sin \left(
\frac{ \mu_0 \tilde{c}}{2} \right)~,
\eeq
where $\sigma_{\rm i}$ are the Pauli matrices and $\tau_{\rm i}=-i\sigma_{
\rm i}/2$ are the basis of the SU(2) Lie algebra and $\mu_0$ is the 
orientated length of the edge with respect to the fiducial metric. The choice
of $\mu_0$ is arbitrary~\cite{ABL} and will now be set to unity. When 
formulated in these terms the evolution equation (the Hamiltonian constraint)
is a discrete equation, with a discreteness scale given by $a_{\rm Pl} = \sqrt{\gamma/6}
l_{\rm Pl}$, where $l_{\rm Pl}$ is the Planck length and $\gamma$ is the Barbero-Immirzi
parameter. This ambiguity comes from the definition of the connection in terms of standard
ADM variables (or to put it another way, an ambiguity in formulating GR in terms of triads
and connections), $A^i_{\rm a} = \Gamma^i_{\rm a} +\gamma K^i_{\rm a}$, where $K^i_{\rm a}$
is the extrinsic curvature one form and $\Gamma^i_{\rm a}$ is the spin connection.
Calculations of black hole entropy\cite{rovelli2,ashtekar_baez_krasnov} fixes this to
be $\gamma \approx .02735$. We will be interested only in effective, continuous equations which
are valid for length scales $a \gg a_{\rm Pl}$. 

Classically the canonical variables $\tilde c, \tilde p$ are related through
\be
\{\tilde c, \tilde p\}=\frac{\kappa\gamma}{3} V_0~,
\ee
where $\kappa\equiv 8\pi G$ and $V_0$ is the volume of the fiducial
cell ${\cal V}$ as measured by the fiducial metric. Defining
\be
p=V_0^{2/3}\tilde p ~~~~ \mbox{and}~~~~ c=V_0^{1/3}\tilde c~,
\ee
with the triad component $p$ determining the {\it physical} volume of the
fiducial cell, and the connection component $c$ determining the 
rate of change of the {\it physical} edge length of the fiducial cell, one
obtains
\be
\{c, p\}=\frac{\kappa\gamma}{3}~,
\ee
independent of the volume of the fiducial cell.

\begin{figure}
 \begin{center}
\input{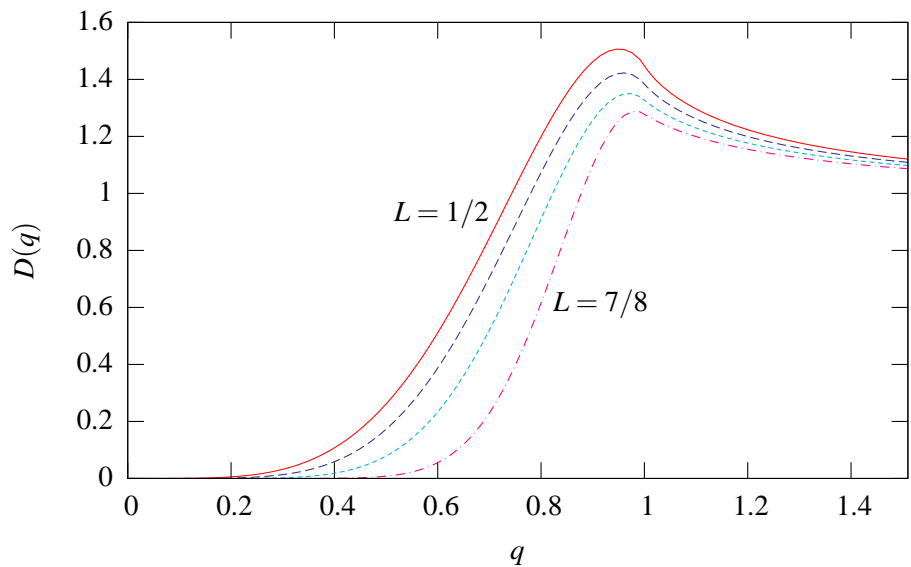}
  \caption{ \label{fig1} The functions $D_{1/2}(q)$,$D_{5/8}$, $D_{3/4}$ and
$D_{7/8}$ are plotted, clearly the quantisation ambiguity, $L$, has no
qualitative effect on the effective equations.}
 \end{center}
\end{figure}

\begin{figure}
 \begin{center}
\input{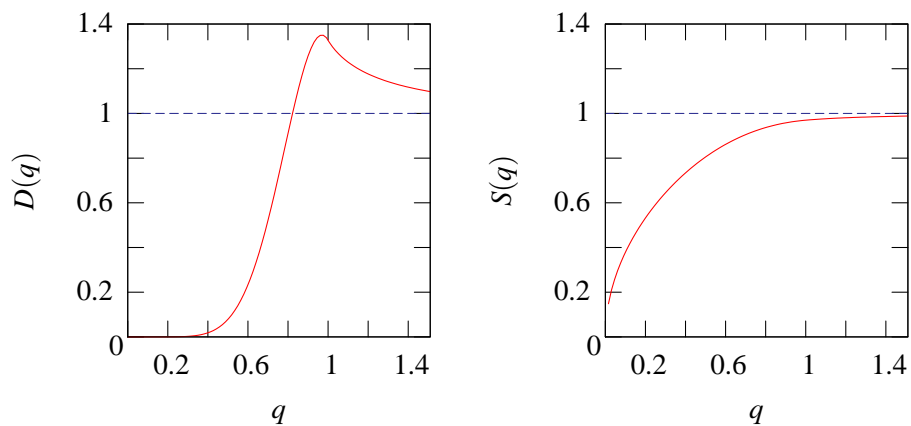}
  \caption{\label{fig2} The function $D(q)$ (left) (with $L=3/4$) is the continuum approximation of the
discrete quantum corrections to the classical inverse volume eigenvalues, whilst $S(q)$ (right)
is the continuum approximation of the quantum corrections to the gravitational part of the
Hamiltonian constraint.}
 \end{center}
\end{figure}

By analogue with the full theory the kinematic Hilbert space
is extended via the Bohr compactification of the real line~\cite{Thiemann1}.
An orthonormal basis for this Hilbert space is $\{ | \mu \rangle \}$,
where
\be
 \langle c | \mu \rangle = e^{i\frac{\mu c}{2}}~.
\ee
The triad operator acts on these basis states as
\be\label{eq:quan_p}
\hat{p}|\mu \rangle=-i\frac{\kappa \gamma \hbar}{3} \frac{\partial}{\partial
 c}|\mu\rangle = \frac{\kappa \gamma \hbar}{6} \mu |\mu\rangle~,
\ee
Clearly the volume operator $\hat{V}\equiv \hat{a}^3=
|\hat{p}|^{3/2}$ is also an eigenstate of this basis,
\be 
\label{eq:vol}
\hat{V}|\mu\rangle = V_\mu|\mu\rangle =  \left(\frac{\kappa \gamma \hbar
 |\mu|}{ 6}\right)^{3/2} |\mu\rangle~. 
\ee
To calculate the eigenvalues of the inverse volume operator, the classical
expression~\cite{Thiemann1},
\be\label{eq:inv}
 \{ c, p^L \} = \frac{\kappa \gamma L}{3} p^{L-1}~,
\ee
where $0<L<1$ is a  quantisation ambiguity, is used. After
quantisation the inverse volume operator acts as~\cite{Vandersloot2005}
\be\label{eq:inv_vol}
 \widehat{V^{-1}} |\mu\rangle = \Biggl( \frac{9}{\kappa\gamma\hbar
L J(J+1)(2J+1)} \sum_{m=-J}^J m V^{2L/3}(\mu+2m\mu_0) \Biggr)
^{3/2(1-L)}|\mu\rangle~.
\ee
The eigenvalue can be approximated by a continuous function~\cite{martinetal}
$d_{J,L} (a)=D_{L}(q)a^{-3}$, where $q=a^2/a_{\star}^2$ and $a_\star^2 = \gamma 
l_{\rm pl}^2 J /3$. The function $D_L (q)$ (see figure~(\ref{fig1}) ) accounts
for the difference between the LQC and the classical inverse volume eigenvalues,
\beq\label{eq:D1}
 D_{L}(q)  &=& \Biggl( \frac{3q^{1-L}}{2L} \Bigl[ \frac{1}{L+2}
\left( \left( q+1\right)^{L+2}-\left| q-1 \right|^{L+2} \right)
\nonumber \\
&&-\frac{q}{1+L} \left(  \left(q+1\right)^{L+1} - {\rm sgn}
\left(q-1\right) \left| q-1\right|^{L+1} \right) \Bigr] \Biggr)^{3/(2-2L)}~.
\eeq
The Hamiltonian constraint is given by~\cite{Vandersloot2005},
\be\label{eq:NSAham_quant}
 ^{(J)} \hat{\cal C}_{\rm g} = \frac{3i \ {\rm sgn}(|p|)}{\kappa^2 
\hbar \gamma^3 J(J+1)(2J+1)} \sum_{ijk} \epsilon ^{ijk} {\rm
 tr} \left( \hat{h}_i \hat{h}_j\hat{h}^{-1}_i\hat{h}^{-1}_j \hat{h}_k
\left[ \hat{h}^{-1}_k,\hat{V} \right]\right)~,
\ee
which can be made self-adjoint simply by symmetrising,
\be\label{eq:SAham_quant}
^{(J)} \hat{\cal H}_{\rm g} = \frac{1}{2} \left( ^{(J)}\hat{\cal C}_{\rm g}
+ ^{(J)} \hat{\cal C}^{\dagger}_{\rm g} \right)~.
\ee
The action of this Hamiltonian constraint on the basis states $|\mu\rangle$
results in a complicated difference equation of order $8J$, however it is well
approximated by the equation~\cite{Vandersloot2005}
\be
 {\cal H} = -3a\dot{a}^2 + s(a) {\cal H}_{\phi} = 0~,
\ee
where $s(a)=a S(a^2/a_{{\rm G} \star}^2)$ is a continuous function that approximates the
corrections to gravity terms, with $a_{{\rm G} \star}$ given by $a_{{\rm G} \star}=\gamma
l_{\rm Pl} J_{\rm G}/3$. Defining $q_{\rm G} = a^2/a^2_{{\rm G}\star}$ the function
is (see figure.~(\ref{fig2})),
\be\label{eq:S(a)}
 S(q_{\rm G}) = \frac{4}{\sqrt{q_{\rm G}}} \left( \frac{1}{10} \left( 
\left( q_{\rm G}+1 \right)^{5/2} + {\rm sgn}\left(q_{\rm G}-1\right)
\left| q_{\rm G}-1\right|^{5/2} \right) -\frac{1}{35} \left( \left( 
q_{\rm G}+1 \right)^{7/2} - \left| q_{\rm G}-1\right|^{7/2}\right)\right)~,
\ee
where $L$ is set to $L=3/4$ from now on, since it has no qualitative effects (see Figure~(\ref{fig1})).
Notice also that $J_{\rm G}$ is not necessarily the same as $J$ and that these {\it effective} equations
are valid only in the continuum era, when $a\gg a_{\rm Pl}$.

Here the matter component we are concerned with is the inflaton, for which we
have~\cite{martinetal},
\be\label{eq:hamiltonian}
 {\cal H} =  -\left(\frac{\dot{a}}{a}\right)^2 + \frac{\kappa S(q_{\rm G})}{3} 
\left[ \frac{1}{2} D^{-(n+1)}\dot{\phi}^2 + D^m V(\phi)\right]=0~,
\ee
where $V(\phi)$ is the inflaton potential and $m,n$ are further quantisation
ambiguities arising due the fact that $\hat{V}\widehat{V^{-1}}$ is not unity\footnote{
In fact it has been shown~\cite{Vandersloot2005} that the effective Hamiltonian should include a 
correction term that induces a bounce at small scales. In what follows, such scales
are not reached and the Hamiltonian given here is a good approximation.} .
Defining $H \equiv \dot{a}/a$, we arrive at the effective, loop quantum cosmological
Friedmann equation,
\be\label{hubble}
 H^2 = \frac{\kappa S(q_{\rm G})}{3} \left[ \frac{1}{2} D^{-(n+1)} \dot{\phi}^2
+ D^m V(\phi)\right]~.
\ee
To simplify the following calculation, $S(q_{\rm G})$ will be set to unity, which
is equivalent to choosing $J_{\rm G}=1/2$. More general choices will be discussed
in at the end of section~(\ref{sec:prob}). Once again it is important to note that these
effective equations are valid only for $a \gg \sqrt{\gamma}l_{\rm Pl}$ and $H^{-1}
\ll \sqrt{\gamma} l_{\rm Pl}$. These limits are crucial in the calculation of the
measure on the space of solutions.

The dynamics of the scalar field are given by the conservation equation~\cite{martinetal},
\be \label{cons}
\ddot \phi+\left[3H-(1+n)\frac{\dot D_l}{D_l}\right]\dot
\phi+D_l^{m+n+1}V'(\phi)=0\ ,
\ee
where $V'\equiv\partial V/\partial\phi$.
\section{Cosmological Measures}
To be able to calculate probabilities, a measure on the space of 
solutions of the theory must be defined. Without a full theory of 
quantum gravity to produce such a measure on the initial conditions
there is a significant ambiguity in how it is defined. Despite
this any measure must satisfy certain basic properties~\cite{GHS}:
\begin{enumerate}
\item it must be positive definite (and finite),
\item it must not depend on the choice of variables used,
\item it must respect all the symmetries of the phase space.
\end{enumerate}
The necessity of the first point is self-evident if the measure is to produce
sensible probabilities. Loosely speaking the second point says that the probability
of a universe should be independent of the time at which we choose to calculate
it, whilst the third forbids the introduction of any ad hoc cutoffs into the 
theory.

The importance of placing a measure on the different possible universes is crucial
to modern cosmology since fundamental theories are no longer expected to provide a
unique cosmological solution. Thus the key question in cosmology is, `how likely is 
our universe?'. It is well known that classical general relativity (and the observed
expansion) predict an initial big bang singularity and it is largely excepted that
the full theory of quantum gravity will overcome this breakdown in predictability. 
Because such candidate theories (e.g. LQG, M-theory etc.) typically differ from classical
physics only at the highest energy scales, it is only through cosmology that they can, in
principle be tested. However any signatures imprinted on the universe by these non-classical 
effects are typically expected to be extremely small and so the prediction that universes
such as our own are a `generic' outcome is usually taken evidence in favour of the
theory. However without a measure on the possible solutions any discussion on how
likely a particular outcome is, at best speculative. 

Inflation is the archetypal example and has produced much debate on whether or not
its onset is fine tuned~\cite{HollandWald,KLM,Coule}. Since the purpose of
introducing an inflationary epoch into the early universe was to eliminate the fine
tuning problems associated with the flatness and horizon problems, it can only be
said to have succeeded if it occurs without similar fine tuning. Without
a fundamental theory that {\it predicts} inflation, the issue of its generality can only
be tackled by introducing a measure on the cosmological histories that contain inflation.
The canonical measure proposed by Gibbons, Hawking and Stewart~\cite{GHS} was shown to
be infinite~\cite{HawkingPage}, however more recently Gibbons and Turok~\cite{GT}
demonstrated that the measure can be regularised and that the results (for certain
quantities) are robust. The measure is derived in the next section, however in the
case of LQC the regularisation procedure is not required and so this won't be discussed.

It is important to note that whilst it has turned out to be difficult to produce
a different measure that satisfy all the criteria given above, there is nothing to
say that it cannot be done. This means that results derived with this measure cannot
be directly compared to results derived using a different measure, although it may
be hoped that certain properties within a theory are largely measure independent.

\section{The Measure}\label{sec:measure}
 The canonical cosmological measure of Ref.~\cite{GHS}
is based on the Hamiltonian structure of a theory and satisfies all the
required criteria (although it is by no means unique~\cite{EKS}) and is the one
used here.

As with all phase spaces, that associated with cosmology has a symplectic form,
\be \label{eq:metric}
\Omega = \sum ^{k}_{i=1} {\rm d}P_{i}\wedge {\rm d} Q^{i}~,
\ee
where $Q_{i}$ and $P_{i}$ are the dynamical degrees of freedom and their conjugate
momenta, respectively.  The $k^{th}$ power of this gives the
volume element of the space.  The Hamiltonian constraint restricts the
physical trajectories to lie on a $(2k-1)$-dimensional surface $M$, of
the full phase space, which is what is known as the multiverse. $M$ also
contains a closed symplectic form $\omega = \sum _{i=1}^{k-1}{\rm d}P_{i}
\wedge{\rm d} Q^{i}$, related to $\Omega$ via,
\be
\Omega = \omega + {\rm d} {\cal H} \wedge {\rm d}t\ \Rightarrow \omega
= \Omega |_{{\cal H} =0}~.
\ee
This measure can be produced from the Hamiltonian given in Eq.~(\ref{eq:hamiltonian}),
which includes the effective quantum gravity corrections to the classical equations.

\begin{figure}
  \begin{center}
      \input{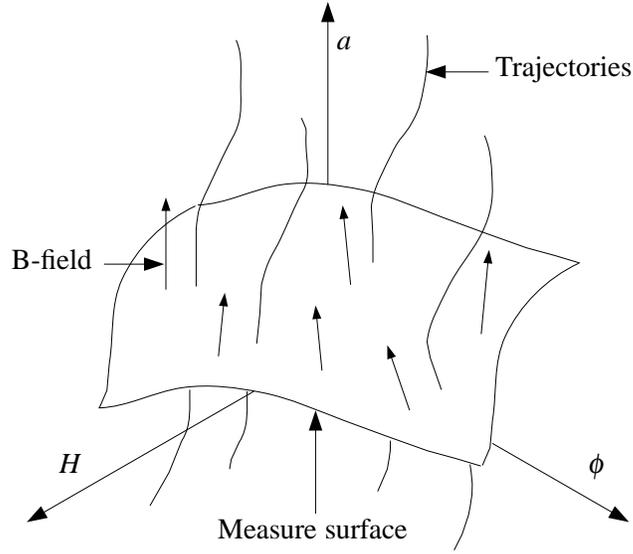}
      \caption{\label{fig:measure} The probability measure is defined
by integrating the ${\bf B}$ field over a constant surface in the
$3$-dimensional phase space produced by using the Hamiltonian
constraint to eliminate one of the dynamical variables.}
  \end{center}
\end{figure}

In this case (a FLRW universe containing only a scalar field) there are only two
canonical variables $(a,\phi)$ and so $k=2$. A divergence-less field ${\bf B}$ can
be defined, given by
\be
B_{a} \equiv \frac{1}{2} \epsilon_{abc} \omega_{bc}~,
\ee
where $\epsilon_{abc}$ is totally anti-symmetric and $a,b,c$ go from $1$ to $3$.
The flow of trajectories in the phase space across a surface is given by ${\bf B}$
evaluated on that surface and this is used to define the measure ${\cal N}$,
\be
 {\cal N} = \int {\bf B} \cdot {\rm d}{\bf S}~,
\ee
where ${\bf S}$ is an open surface crossing each trajectory only once.
The scheme is depicted in Figure~(\ref{fig:measure}).

By construction $\bf{B}$ is divergence-less and so a vector potential can be
defined ${\rm d}\bf{A} = \bf{B}$. Then using Stoke's theorem the measure becomes,
\be
{\cal N} = \oint {\bf A} \cdot {\rm d}{\bf l}~,
\ee
where ${\bf l}= \partial {\bf S}$ is the boundary of ${\bf S}$. This 
ensures that the measure depends only on the flux of trajectories through ${\bf l}$
and is independent of the (topologically equivalent) surface chosen.

\section{Total volume of phase space}
Once a measure on the phase space has been defined the probability of a 
particular set of trajectories can be calculated as the ratio of the 
measure of those trajectories to the measure of the total phase space.
In the classical case this is not well defined~\cite{HawkingPage} due
to a divergence in the total phase space measure, unless a small curvature
cut-off is introduced~\cite{GT}. The effective equations of LQC
remove this divergence, since there is a minimum scale that can be
probed within this formalism, $a \gg \sqrt{\gamma}l_{\rm Pl}$. However
there is a further possible source of divergence in the classical
case that is not removed by the LQC corrections, that restricts the form
of inflationary potential that can be used.

Here a sketch of calculation is given, the details being given in~\cite{us}.
The measure of the total phase space, defined by Eqs.~(\ref{eq:hamiltonian}), (\ref{cons})
can be calculated using the momenta associated with the scale factor and the scalar field
$\phi$,
\be
P_{a}=-6a^{2}H ,\
P_{\phi}= a^{3} D_{l}^{-(n+1)} \dot{\phi}~.
\ee
In terms of $q=a^{2}/a_{\star}^{2}$ and using Eq.~(\ref{hubble}) with $S(q_{\rm G})=1$,
the measure evaluated on a $q=q_{\rm s}$ surface, where
$q_{\rm s}$ is a constant gives,
\begin{equation}
\label{N}
    \mathcal{N} = - \frac{3}{4\pi G} \int \int
\frac{Ha_{\star}^{3}q^{\frac{3}{2}}_{\rm s} D_{\rm s}^{-\left(
\frac{n+1}{2}\right) } }{\sqrt{\frac{3H^{2}}{4\pi G} - 2D_{\rm s}^{m}
V }} {\rm d}H {\rm d}\phi~,
\end{equation}
where from now on the $l$ label is dropped and the notation
$f(a_{\rm s})=f_{\rm s}$ is used.

The $H$ integral can be performed to give,
\begin{equation} \label{N2}
    \mathcal{N}=- a_{\star}^{3}q^{\frac{3}{2}}_{\rm s}D_{\rm
    s}^{-\frac{n+1}{2}}  \int_{\phi_{\rm i}}^{\phi_{\rm f}}
    \sqrt{ \frac{3}{4\pi\gamma l_{\rm Pl}^{4} }
    -2D_{\rm s}^m V(\phi)}{\rm d}\phi\ ,
\end{equation}
where $[\phi_{\rm i},\phi_{\rm f}]$ is the range on $\phi$ which
ensures ${\cal N}$ is real.

It is clear Eq.~(\ref{N2}) is by not always convergent, as was assumed
in Ref.~\cite{GT}, in the classical ($D\rightarrow 1$) case for potentials
with only one minima. However it does converge for many physically
interesting potentials (e.g. single minima potentials that are unbounded above).
In the following sections specific potentials will be used to calculate probabilities.
For the moment however all that is required is that the potential is such that Eq.~(\ref{N2})
converges.

\section{Probability}\label{sec:prob}
To calculate the probability of inflation, the measure of only the 
inflationary trajectories is needed i.e. 
\be
 {\cal M} \equiv {\cal N}\big|_{\rm inflation}~.
\ee

Using Stoke's theorem for the path $H=H_{\rm s}={\rm const.}$ on Eq.~(\ref{N})
restricted to only inflationary trajectories, we have
\be \label{M}
    {\cal M}=- \oint \mid P_\phi\mid {\rm d}\phi{\Big |}_{\rm inflation}\ ,
\ee
which is positive as inflation runs from higher to lower values of the scalar field $\phi$.
Details of how this can be calculated and how the calculation differs from the classical
case can be found in Ref.~\cite{us}. The result is
\be\label{M2}
    {\cal M} = \frac{a_{\star} ^{3}q^{\frac{3}{2}}_{\rm s} }{4\pi G}
D_{\rm s} ^{-\left( \frac{n-m+1}{2} \right) } \delta \epsilon_{\rm i}
 \exp \left( -3 \Big| 1 - (n-m+1)\frac{q_{\rm
s}}{3D_{\rm s}} \left( \frac{\partial D}{\partial q} \Big|_{\rm s}
\right) \Big| N \right)~,
\ee
where
\be \label{de}
 \delta \epsilon_{\rm i} \approx \frac{1}{12 A^{3}_{\rm s}}
 \sqrt{\frac{V(\phi_{\rm i})}{24\pi G}}
 \left( \frac{1}{V(\phi_{\rm i})}\ \frac{ \partial V(\phi)
 }{\partial \phi} \Big|_{\rm i} \right)^{2}~,
\ee
\be
 A\equiv \Bigl| 1 - \left( n-m+1\right) \frac{q}{3D} \frac{\partial D}{
\partial q} \Bigr| D^{-(n+1)/2}~,
\ee
and the constant $q=q_s$ surface on which the measure is being calculated is
taken to be at the end on inflation i.e. at the breakdown of the slow-roll
conditions.
To be able to evaluate this a potential must be given and here potentials
of the form,
\be
\label{powerlaw}
V(\phi)=\frac{\mu^4}{2\alpha!}\ \left(\frac{\phi}{\mu}\right)^{2\alpha}~,
\ee
are considered. Notice that these belong to the class of potentials that
make Eq.~(\ref{N2}) finite and so well defined probabilities can be produced.
Using the slow-roll approximation (and the assumption that $\phi_{\rm f} \ll 
\phi_{\rm i}$) with this class of potentials, Eq.~(\ref{de}), can be shown to be,
\be
\label{deltaepsiloni}
\delta\epsilon_{\rm i}\approx \frac{l_{\rm
Pl}^{-1}}{\sqrt{2\alpha!}}\frac{\alpha^2}{3 A_{\rm
s}^3\sqrt{24\pi}}\left(\frac{\mu}{\phi_{\rm i}}\right)^{2-\alpha}\ ,
\ee
where
\be
\phi_{\rm i}\approx \left(\frac{\alpha}{4\pi
G}\int^{a_{\rm s}}_{a_{\rm i}}\frac{D^{n+1}}
{[\usual]}\frac{{\rm d}a}{a}\right)^{1/2}\ .
\ee
Notice that for $\alpha=1$ and $D=1$ we recover the standard GR result
$\phi_{\rm i}\approx \sqrt{N/(4\pi G)}$.

Finally, from Eqs.~(\ref{N2}), (\ref{M2}) and (\ref{deltaepsiloni}), the
probability ${\cal P}(N)$ of having  $N$ e-folds of slow-roll inflation is
\be \label{explicitprob}
{\cal P}(N)\approx \beta^2 \left(\frac{\mu}{\phi_{\rm
i}}\right)^{2-\alpha}\left(l_{\rm
Pl}\mu\right)^{\frac{2-\alpha}{\alpha}}
\exp \left(-3 \Big| 1 -
\left( n -m +1 \right) \frac{q_{\rm s}}{3D} \left({\partial D
\over\partial q}\Big|_{\rm s} \right) \Big| N \right)~,
\ee
where
\be\label{prob}
 \beta^2=\frac{\alpha^3}{144}
\left[\frac{2}{3\pi(2\alpha!)}\right]^{\frac{\alpha+1}{2\alpha}}
2^{\frac{\alpha+2}{2\alpha}}\pi^{\frac{\alpha-1}{\alpha}}
\gamma^{\frac{\alpha+1}{2\alpha}}
\frac{\Gamma(\frac{3\alpha+1}{2\alpha})}{\Gamma({1\over 2\alpha})}
D_s^{m\frac{\alpha+1}{2\alpha}} A_s^{-3}~.
\ee
This changes qualitatively for renormalisable ($\alpha\leq 2$) and
non-renormalisable ($\alpha>2$) potentials and here we concentrate only
on renormalisable potentials~\cite{pea} i.e. $\alpha=1,2$.

The above calculation can be repeated using $S(q_{\rm G}) \neq 1$ to
give
\be
{\cal P}(N)\approx \beta^2 \left(\frac{\mu}{\phi_{\rm
i}}\right)^{2-\alpha}\left(l_{\rm
Pl}\mu\right)^{\frac{2-\alpha}{\alpha}} \biggl[S\biggl({a^2\over
a_G^2}\biggr)\biggr]^{\frac{\alpha+4}{4\alpha}}
\times \exp \left(-3 \Big| 1 - \left( n -m +1 \right) \frac{q_{\rm
s}}{3D} \left({\partial D \over\partial q}\Big|_{\rm s} \right) \Big|
N \right)~,
\ee
where now, \be \phi_{\rm i}\approx \left(\frac{\alpha}{4\pi
G}\int^{a_{\rm s}}_{a_{\rm i}}\frac{D^{n+1}} {S\biggl({a^2\over
a_G^2}\biggr)\biggl[\usual\biggr]}\frac{{\rm d}a}{a}\right)^{1/2}\ .
\ee
Since $S(q_{\rm G})<1$ it is clear that choosing $j_{\rm G} \neq 1/2$
(i.e. $S \neq 1$) slightly reduces the probability of inflation.
However, for $a_{\rm i}>a_{{\rm G}}$, $S(q_{\rm G})$ is, well approximated
by 1, thus the conclusions for the $j_{\rm
G}=1/2$ case remain qualitatively unchanged. Therefore in the following
we shall only consider the $j_{\rm G}=1/2$ fundamental representation case.

\section{Evaluating the probability}
Whilst Eq.~(\ref{prob}) looks much more complicated than the classical result
given in Ref.~\cite{GT}, it can be simplified significantly.
In order to have sufficient inflation to solve the standard cosmological
problems, approximately $60$ efoldings is required i.e.
\be
\label{nec1}
e^{60}\approx \frac{a_{\rm s}}{a_{\rm i}}< \frac{a_{\rm s}}{a_{\rm
Pl}}\ ,
\ee
where $a_{\rm Pl}$ is the minimal scale which can be probed by the 
effective equations of LQC used here. The scale at which quantum
effects become significant, $a_\star$, is governed by $j$. This ambiguity
has been restricted by particle physics experiments to be $j<10^{20}$, from
which it is easy to see that $a_{\rm s} \gg a_\star$ (i.e. inflation ends
well into the classical epoch), allowing functions evaluated on $a_{\rm s}$
to be expanded in the $q \gg 1$ limit.

The first term in the probability, Eq.~(\ref{prob}) that is of interest is
the exponential suppression factor
\be
\exp\left(-3\left[\usual\right]\Big|_{a_{\rm s}} N\right)~.
\ee
It seems possible to overcome the classical exponential suppression simply by 
choosing $n,m$ and $a_{\rm s}$ so that $[ \usual ]\Big|_{a_{\rm s}}\sim 
{\cal O}(0)$. However the probability is also proportional to 
\be
\beta^2\propto \left[\left(\usual\right)\Big|_{a_{\rm s}}\right]^{3}~.
\ee
Therefore the highest probability is found when $[ \usual ]\Big|_{a_{\rm s}}\sim 
{\cal O}(1)$. Thus the exponential suppression present in the classical ($D\rightarrow
1$) case found in Ref.~(\cite{GT}) is not removed.

The most interesting term is $\mu/\phi_{\rm i}$, since $\phi_{\rm i}$
contains an integral over all scales. Thus the behaviour of $[\usual ]^{-1}$ throughout
inflation must be investigated. In order to significantly increase the probability,
there cannot be a zero of $[\usual ]$ at any point during inflation, which implies
that $a_{\rm i}$ has to be larger than the scales of any such zeros. It can be shown~\cite{us}
that this factor does not significantly increase the probability (relative to the exponential
suppression), for values $N\gtrsim 22.5$ (the $22.5$ arises because of the experimental
limits on $j$ and is only relevant for a very small range of $n-m+1$).

Finally the probability contains the factor $\beta^2\propto D_{s}^{\frac{m
(\alpha+1)-\alpha(n+1)}{2\alpha}}$. Since $D>1$ at the end of inflation, for
sufficiently large powers it is possible for this term to become large. However
it can only overcome the exponential suppression for ambiguity parameters
that satisfy
 \be
 \frac{m(\alpha+1)-\alpha(n+1)}{2\alpha}\gtrsim 10^{110}\ .
\ee
Clearly this cannot be considered a {\it natural} choice.

\section{Conclusion and discussion}
Cosmological inflation can only be said to solve the many fine-tuning problems 
associated with the hot big bang model if it is a generic outcome of
a theory. This problem has been faced in the past~\cite{piran, calzettamairi}
and has recently been highlighted by Ref.~\cite{GT}. In that study
it was shown that, with a particular choice of measure, the probability
of having sufficient inflation is exponentially suppressed by the 
number of efolds. This is an important result and its validity and 
generality should be proven. One major difficulty with the calculation
of Ref.~\cite{GT} is that classical physics was assumed to be valid
at all scales during inflation, however it is well known that quantum
corrections can no longer be neglected during the early phases of 
inflation. Here (and in detail in Ref.~\cite{us}) it has been shown how to 
include these quantum corrections within the formalism of effective
continuum loop quantum cosmology.

Applying the canonical measure proposed by Ref.~\cite{GHS} to the
effective equations of loop quantum cosmology allowed the probability
of having sufficient, single field slow-roll inflation to be calculated.
The resulting probability was then examined for the phenomenologically
important $V\sim \phi^2$ and $V\sim \phi^4$ inflationary scenarios.
It was shown that this probability is exponentially suppressed for all
but the most extreme values of ambiguity parameters. In particular
the values $m=n=0$ that are typically used in the literature do not
alleviate the exponential suppression.
It is known~\cite{mb} that LQC can produce an era of {\it super-inflation}, in
which the scalar field is driven up its potential. Such an epoch would not
satisfy the slow-roll approximation used here and so would not be counted in
this probability. However in this era perturbation theory is
unstable~\cite{Tsujikawa:2003vr} and so to produce the observed CMB anisotropies
$60$ {\it subsequent} efoldings of standard inflation are required and it is this
that our probability considers.

Our findings do not imply that inflation itself is improbable, what they do
show is that, at least in the case of the semi-classical
regime of loop quantum cosmology and therefore of general relativity,
inflation is not probable {\it with this particular measure}. Whenever 
cosmological probabilities are discussed a measure needs to be defined
and, as has been shown here, this can lead to vastly differing results.
Previously analysis on the likelihood of inflation (see for example~\cite{KLM})
have used the {\it prior} that all initial conditions are equally likely, whilst here
a late time equivalence of universes that resemble our own has been taken. In the 
former case inflation was shown to be an attractor solution, whilst in the 
latter it is not. Which prior you choose to assume is a matter of taste,
however what is clear is that any conclusions drawn will be highly sensitive
to that choice. The crucial point then is that inflation may not be as
likely as previously assumed.

To produce a definitive result one has to
address inflation in full quantum gravity, or in a string theory context, so as
to fully understand the initial conditions of the universe.
Without such a fundamental treatment results on the generality of 
inflation will invariably be dependent on the choice of measure. 

\section*{Acknowledgements}
This work has been supported by the European Science Foundation network
programme "Quantum Geometry and Quantum Gravity".

\end{document}